\documentclass[prb,preprint]{revtex4-1} 

\usepackage{amsmath}  
\usepackage{amsfonts} 
\usepackage{graphicx} 
\usepackage{subfig}
\usepackage{hyperref}
\usepackage{multirow}

\begin{document}


\title{Numerical Simulation Study of Neutron-Proton Scattering using Phase Function Method}

\author{Shikha Awasthi}
\author{Anil Khachi}
\author{Lalit Kumar}
\author{O.S.K.S. Sastri}
\email{sastri.osks@hpcu.ac.in} 
\affiliation{Department of Physics and Astronomical Science, Central University of Himachal Pradesh, Dharamshala - 176215, Bharat (India)}


\date{\today}

\begin{abstract}
In this article, we propose a numerical approach to solve quantum mechanical scattering problems, using phase function method, by considering neutron-proton interaction as an example. The nonlinear phase equation, obtained from time-independent Schr$\ddot{\text{o}}$dinger equation, is solved using Runge-Kutta method for obtaining S-wave scattering phase shifts for neutron-proton interaction modeled using Yukawa and Malfliet-Tjon potentials. While scattering phase shifts of S-states using Yukawa match with experimental data for only lower energies of $50$~MeV, Malfliet-Tjon potential with repulsive term gives very good accuracy for all available energies up to $350$~MeV. Utilizing these S-wave scattering phase shifts, low energy scattering parameters and total S-wave cross section have been calculated and found to be consistent with experimental results. This simulation methodology can be easily extended to study scattering phenomenon using phase wave analysis approach in the realms of atomic, molecular and nuclear physics. 

\end{abstract}

\maketitle 

\section{Introduction}
The wavefunction obtained by solving the time-independent Schr$\ddot{\text{o}}$dinger equation for various models of interaction potentials is central to understanding quantum mechanical systems. Problems involving penetration through a rectangular barrier, and bound and scattering states of finite square well are routinely performed in an undergraduate quantum mechanics course. Gamow's theory of alpha ($\alpha$) particle tunneling based on extension of barrier penetration and explanation of bound state of deuteron ($d$) and neutron-proton (\textit{np}) scattering cross section utilizing square well are part of nuclear physics course at undergraduate level. Even though Yukawa's meson exchange theory\cite{Yukawa} to explain \textit{np}-scattering is discussed at undergraduate level, corresponding time-independent Schr$\ddot{\text{o}}$dinger equation for Yukawa potential is not solved. This is mainly due to non-availability of analytical solution, till recently\cite{quantYuk}. Thus, most textbooks on nuclear physics\cite{Krane,Wong,Hans} still rely on square well potential for obtaining binding energy of deuteron as well as discussing \textit{np} scattering cross section. So, there is a need to introduce a numerical technique to solve Yukawa potential to provide a better understanding of \textit{np}-scattering. This would equip students with required skills to solve two-body scattering problems in atomic, nuclear, and particle physics. \\ 
The main objective of this paper is to present a simple derivation of phase equation for S-wave (i.e. $\ell=0$) and solve it numerically using Runge-Kutta method to obtain scattering phase shifts for neutron-proton \textit{np} - interaction by choosing Yukawa potential and its modified form called Malfliet-Tjon (MT)\cite{MT} potential.\\
Theoretically, scattering cross section data are obtained from scattering phase shifts (SPS) by using either effective range theory or phase shift analysis. The later approach involves determination of scattering phase shifts that arise due to scattering of incoming projectile with interaction potential of target nucleus. Theoretical modeling involves, proposing a mathematical function to represent interaction potential based on understanding of underlying physical phenomena and then solving the radial time-independent Schr$\ddot{\text{o}}$dinger equation to obtain the wavefunction. Mostly, scattering phase shifts are deduced from matching the wave function within interaction region with that of asymptotic region in which interaction ceases to exist\cite{rmatrix,smatrix,jost}. This wave function approach to determining scattering phase shifts involves solving the radial time-independent Schr$\ddot{\text{o}}$dinger equation numerically and is discussed in advanced computational physics book\cite{Thijssen}, which is beyond the reach of many undergraduate physics students.\\
Experimentally, scattering cross section data are available at different lab energies of incoming projectile, from which scattering phase shifts are deduced. The scattering phase shifts data for nucleon-nucleon [neutron-proton $(np)$, neutron-neutron $(nn)$ and proton-proton $(pp)$]\cite{Anil}, nucleon-nucleus [neutron-deuteron $(nd)$, proton-deuteron $(pd)$, neutron-alpha $(n\alpha)$, proton-alpha $(p\alpha)$]\cite{Triton, Lalit} and nucleus-nucleus [$\alpha\alpha$]\cite{Alpha} systems are available in literature. \\ 
An alternative approach to determine scattering phase shifts is variable phase approach, originally proposed by Morse\cite{Morse}, and later came to be known as phase function method (PFM)\cite{Babikov, Calogero, Kahn}. This method has been utilized for obtaining scattering phase shifts for \textit{np}-interaction with reasonable success\cite{AnilChitkara, Zhaba, Laha}. In this approach, the time-independent Schr$\ddot{\text{o}}$dinger equation is transformed into a first order non-linear Ricatti-type equation that directly deals with phase shifts for different $\ell$ values and different energies, without need for wavefunction like other methods. Thus, phase function method is an easy alternative to traditional methods like r-matrix method\cite{rmatrix}, s-matrix method\cite{smatrix} or jost function method\cite{jost}, etc.\\ 
In this work, we have solved phase equation numerically by choosing Runge-Kutta $5^{th}$ order\cite{rk5} (RK-5) method to obtain scattering phase shifts for $^3S_1$ and $^1S_0$ states in \textit{np} interaction using Yukawa and Malfliet-Tjon potential. By providing best model parameters for both these potentials, RK-5 method can be easily implemented using worksheet environment such as Gnumeric, Excel or LibreOffice-Calc for determining scattering phase shifts at different energies. This would be within the reach of undergraduate physics students. The obtained scattering phase shifts are then utilized to determine total scattering cross section at various energies and scattering parameters for both singlet and triplet states. \\
The paper is structured based on simulation methodology\cite{Aditi1} consisting of following four stages:
\begin{enumerate}
\item \textit{Modeling physical system}\cite{AJP}\\ In next section, we will describe scattering process in detail and formulate mathematical model in terms of phase equation. Then, numerical solution is developed in three steps.
\item \textit{Preparation of system} by choosing appropriate units, region of interest and numerical technique.
\item \textit{Implementation of numerical method} in a computer.\\
These two stages are briefly touched upon in Section III. This is followed by
\item \textit{Simulation of results and discussion} in Section IV\\
and conclusions are given in last section.
\end{enumerate} 

\section{Modeling physical system: Description and Formulation}
\noindent{The process of scattering of a neutron with energy $E_{\ell ab}$ and a proton, which is at rest in lab frame\cite{Sarsour}, is represented with position vectors $\vec{r}_n$ and $\vec{r}_p$.} Their masses are $m_n$ and $m_p$ respectively. This two body system is reduced to a one body system by transformation to center of mass frame. In this process, origin is shifted to center of mass co-ordinate $\vec{R}_{cm}$ and two particles are replaced by a single particle with reduced mass $\mu_D = (m_n m_p)/(m_n + m_p)$ which has a position vector $\vec{r}$, that represents relative distance between neutron and proton. The projectile energy in lab frame, $E_{\ell ab}$, would be related to centre-of-mass energy $E_{cm}$ using standard relation\cite{Zettli, Paneru}:
\begin{equation}
E_{cm} = \Big(\frac{m_T}{m_P+m_T}\Big)E_{\ell ab}~~=~~ \Big(\frac{m_p}{m_n+m_p}\Big)E_{\ell ab}
\end{equation}
Where, $m_P$ is the mass of projectile (neutron) and $m_T$ is the mass of target (proton). For simplicity, we drop the subscript henceforth and write center of mass energy as $E$.
The ideal choice for reference system would be spherical polar co-ordinates, $\vec{r} = (r,\theta,\phi)$, as potential has central force characteristics. The state of system, for $\ell=0$ is described by its wavefunction $\psi(\vec{r},t)$, and is obtained by solving the radial time-independent Schr$\ddot{\text{o}}$dinger equation, given by   
\begin{equation}
\frac{d^2u_{0}(r)}{dr^2}+\frac{2 \mu}{\hbar^2}\big[E-V(r)\big]u_{0}(r) = 0
\label{rTISE} 
\end{equation}
The wavefunction must satisfy $u_0(0) = 0$ at r = 0. Further, at a distance $r_0$ beyond which V(r) is zero, wavefunction and its derivative both need to be continuous. That is, choosing $u_a(r)$ to be asymptotic solution of Eq. \ref{rTISE} for $r > r_0$, we must have
\begin{equation}
u_0(r)\big|_{r = r_0} = u_a(r)\big|_{r = r_0} 
\label{wfc}
\end{equation}
similarly 
\begin{equation}
\frac{du_0(r)}{dr}\bigg|_{r = r_0} = \frac{du_a(r)}{dr}\bigg|_{r = r_0} 
\label{wfdc}
\end{equation}
These two conditions are combined together into a single equation by considering logarithmic derivative to satisfy boundary condition, obtained as 
\begin{equation}
\frac{1}{u_0(r)}\frac{du_0(r)}{dr}\bigg|_{r = r_0} = \frac{1}{u_a(r)}\frac{du_a(r)}{dr}\bigg|_{r = r_0} 
\label{logder}
\end{equation}
\subsection{Concept of Phase-shift:}
\begin{figure}[hbtp]
\centering
\includegraphics[scale=0.7]{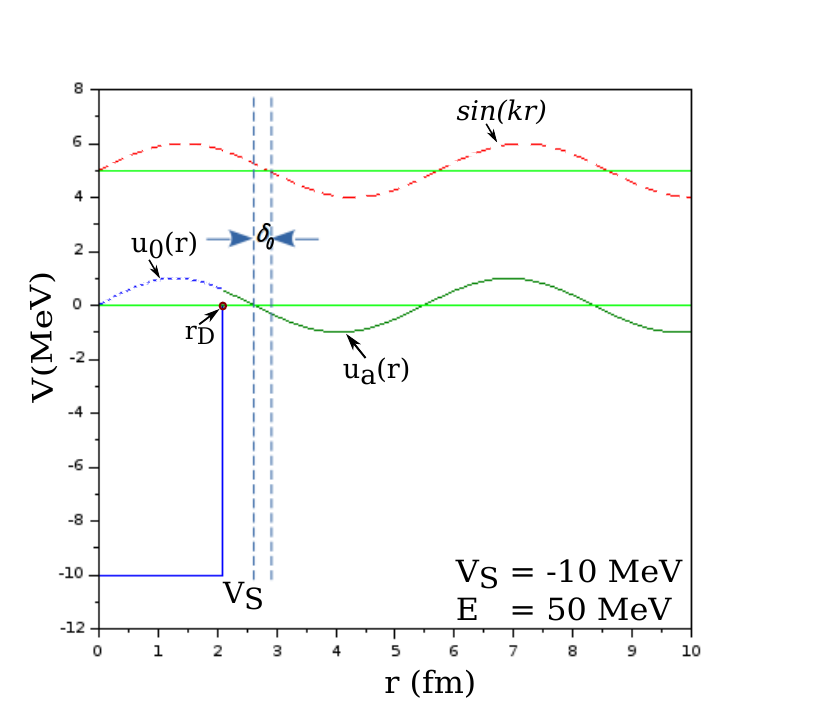}
\caption{The square-well potential $V$~(MeV) is plotted w.r.t distance `r'. The singlet state wavefunction $u_0(r)$ within the well is matched with asymptotic solution $u_a(r)$ outside it at $r_D$.}
\label{phase-shift}
\end{figure}

\noindent{Phase shift techniques are incredibly helpful when analyzing scattering, including nucleon-nucleon scattering. The usefulness of approach to obtain phase shift using wavefunction is applied to $np$-scattering on a square-well potential in Krane\cite{Krane}. Here, we are including spin-spin interaction\cite{MIT notes} and numerically obtain phase shifts for $^1S_0$ state of Deuteron.
This lays foundation for deriving phase equation which is central to phase function method.}\\
The width of square well for deuteron system is known to be $r_0 = r_D = 2.1$~fm\cite{Bertulani}, radius of $np$ - system. The depth of square well can be determined to match binding energy of deuteron\cite{Krane}. The triplet ground state has energy $E = -2.2$~MeV and virtual singlet state has energy $E = 77$~keV. Due to spin-dependence, depths $V_T$ of triplet ($^3S_1$) state  and $V_S$ for singlet ($^1S_0$) state are determined to be respectively $-32.5$~MeV and $-10$~MeV\cite{MIT notes}.\\
For $E > 0$ singlet state, solution within region of well would be given by
\begin{equation}
u_0(r) = A~\sin(k_0r) + B~\cos(k_0r) = A_0 \sin(k_0r+\phi_0)
\end{equation}
where $k_0 = \sqrt{\frac{2\mu_D(E - V_S)}{\hbar^2}}$.\\
The boundary condition at r=0 gives $\phi_0=0$.\\ 
Similarly, for asymptotic solution outside the well, where there is no interaction, one obtains 
\begin{equation}
u_a(r) = A_a~\sin(k_ar + \delta_0)
\label{ua}
\end{equation} 
where 
$k_a = \sqrt{\frac{2\mu_D E}{\hbar^2}}$.\\
The logarithmic derivatives of these two wavefunctions are matched at boundary $r = r_D$, to obtain
\begin{equation}
k_1~\cot(k_1r_D) = k~\cot(kr_D + \delta_0)
\end{equation}
This equation gives us value of $\delta$. These two solutions $u_0(r)$ and $u_a(r)$ are plotted in Fig.\ref{phase-shift} for $E = 50$~MeV. The free particle solution $u(r)= \sin(kr)$ is also plotted in upper part to visually indicate phase shift accrued due to interaction with square well potential. This way of obtaining numerically the phase shift associated with scattering state of deuteron is seldom done in any textbook, such an activity would enhance learning of the important concept.
\subsection{Model Interaction Potentials:}
\noindent{The interaction between neutron and proton is originally modeled successfully by Yukawa\cite{Yukawa} as}
\begin{equation}
V_{Y}(r)= -V_A\Big(\frac{e^{-\mu_A r}}{r}\Big)
\label{MT}
\end{equation}
where $V_A$ is strength of interaction in MeV and $\mu_A~fm^{-1}$ is screening parameter which reflects range of interaction. We will be initially solving phase equation for this potential for various lab energies to show its relevance for low energies. Then, for including role of higher energies, a repulsive part of similar form is added, as proposed by Malfliet and Tjon\cite{MT}, given by:
\begin{equation}
V_{MT}(r)= -V_A\Big(\frac{e^{-\mu_A r}}{r}\Big)+ V_R\Big(\frac{e^{-\mu_R r}}{r}\Big)
\label{MT}
\end{equation}
where, they chose $\mu_R = 2\mu_A$. We will refer to it as Malfliet-Tjon (MT) potential and has three parameters.\\
Instead of numerically solving for wavefunction, we introduce phase function method which results in scattering phase shifts directly from interaction potential.
\subsection{Phase function method:}
\begin{figure}[h!]
    \centering
     \includegraphics[width=3.9in, height=3.4in]{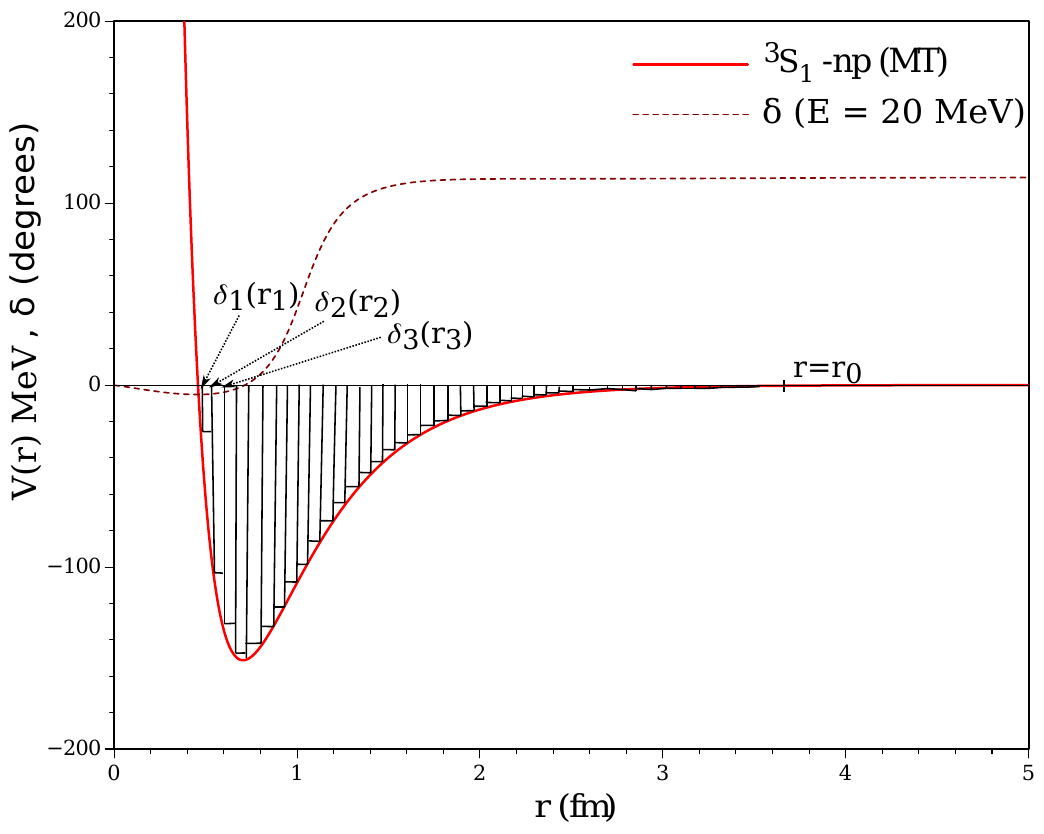}
    \caption{MT potential for $^3S_1$ state, is shown as series of finite rectangular wells. The dotted line shows build up of SPS $\delta_i(r_i)$ at lab energy $E = 20$~MeV with distance ($r$).}
    \label{effect}
\end{figure}

\noindent{\textbf{Derivation of phase equation}}\\
The transformation of second order time-independent Schr$\ddot{\text{o}}$dinger equation into a Ricatti type first order non-linear differential equation i.e phase equation for $\ell=0$, was initially given by Morse and Allis\cite{Morse2}. It was later generalised for higher partial waves by Babikov\cite{Babikov} and Calegero\cite{Calogero}. Here, we present the derivation for $\ell=0$ case\cite{Morse2} in a pedagogical manner.\\  
We now introduce the following visual explanation for the first time. Consider a general potential as shown in Fig. \ref{effect}, to be a combination of extremely small square wells, of differing depths, $V_i(r_i)$.\\ 
The wavefunction for the $i^{th}$ well between $r_i$ to $r_{i+1}$ would be $u_i(r)=A_i \sin(k_ir+\phi_i)$, where $k_i=\sqrt{2\mu(E - V_i(r_i))/\hbar^2}$. Here, $\phi_i$ is determined by matching the boundary conditions at $r_i$, by considering the wavefunctions in the square wells from $r_{i-1}$ to $r_i$ and $r_i$ to $r_{i+1}$. Remember, that $\phi_0=0$ for the first well between $r_0$ and $r_1$. Then, $\delta_i$ is obtained by matching wavefunction at $r_{i+1}$ to asymptotic solution in Eq. \ref{ua}.
So, one would have
\begin{equation}
\frac{1}{u_i(r)}\frac{du_i(r)}{dr}\bigg|_{r = r_{i+1}} = \frac{1}{u_a(r)}\frac{du_a(r)}{dr} \bigg|_{r = r_{i+1}}
\label{boundary}
\end{equation}
By substituting asymptotic solution from Eq. \ref{ua} and defining a function $Z_i(r_{i+1})$ as
\begin{equation}
Z_i(r_{i+1}) = \frac{1}{u_i(r)}\frac{du_i(r)}{dr}\bigg|_{r = r_{i+1}} = k_a \cot(k_a r_{i+1}+\delta_i(r_{i+1}))
\label{Zi}
\end{equation}
As width of wells tends to $0$, the approach moves from discrete to continuous. Then $u_i(r)$ is replaced by $u(r)=A\sin(kr+\phi)$, where $k=\sqrt{2\mu(E - V(r))/\hbar^2}$ and $Z_i(r_{i+1})$ becomes 
\begin{equation}
Z(r) = \frac{1}{u(r)}\frac{du(r)}{dr} = k_a \cot(k_a r+\delta(r))
\label{logdergen}
\end{equation}
Derivative of Z(r), using Eq. \ref{logdergen} is
\begin{equation}
\frac{dZ(r)}{dr} = -\frac{(k_a^2+k_a\frac{d\delta}{dr})}{\sin^2(k_ar+\delta)}
\label{Ader}
\end{equation}
Now, differentiating $Z(r)$, by using first part of Eq. \ref{logdergen}, i.e within the potential region, one obtains
\begin{equation}
\frac{dZ(r)}{dr}=\frac{d}{dr}\bigg(\frac{1}{u(r)}\frac{du(r)}{dr}\bigg) = \frac{1}{u(r)}\frac{d^2u(r)}{dr^2} - \frac{1}{u^2(r)}\bigg(\frac{du(r)}{dr}\bigg)^2
\end{equation}
From which, we obtain 
\begin{equation}
\frac{1}{u(r)}\frac{d^2u(r)}{dr^2} = \frac{dZ}{dr} + Z^2(r) 
\end{equation}
\underline{\textit{Transforming radial time-independent Schr$\ddot{\text{o}}$dinger equation into phase equation:}}\\
Dividing equation Eq. \ref{rTISE} by $u(r)$, we get
\begin{equation}
\frac{1}{u(r)}\frac{d^2u(r)}{dr^2} + \frac{2\mu}{\hbar^2}(E-V(r)) = 0     
\end{equation}
In terms of $Z(r)$, it is written as
\begin{equation}
\frac{dZ(r)}{dr} + Z^2(r) = \frac{2\mu}{\hbar^2}\big(V(r)-E\big)
\label{Aeqn}
\end{equation}
On substituting from Eqs. \ref{logdergen} and \ref{Ader}, we have
\begin{equation}
-\frac{(k_a^2+k_a\frac{d\delta}{dr})}{\sin^2(k_ar+\delta)} + k_a^2\frac{\cos^2(k_ar+\delta)}{\sin^2(k_ar+\delta)}=\frac{2\mu}{\hbar^2}(V(r)-E)
\end{equation}
Using relation $\cos^2\theta=1-\sin^2\theta$ and $k_a^2 = \frac{2\mu E}{\hbar^2}$, it gets simplified to result in following phase equation 
\begin{equation}
\boxed{
\frac{d\delta(r)}{dr} = -\frac{2\mu}{\hbar^2}\frac{V(r)}{k_a}~\sin^2(k_ar+\delta(r))}
\label{pheqn}
\end{equation}
This is a non-linear first order differential equation of Ricatti type, with initial condition $\delta$($r = 0$)$=0$. Eq. \ref{pheqn} can not be solved using any analytical techniques and hence we resort to numerical approach. One can obtain the wavefunctions from phase shifts\cite{Calogero}, but this is not necessary for determination of experimental scattering parameters or cross section, and hence is not attempted in this work.
\section{Numerical Solution}
\subsection{Preparation of System:}
\noindent{\textbf{\underline{Choice of units:}}} 
In nuclear physics, scale of energies are in MeV and that of distances in fm. Converting J-m to MeV-fm, value of $\hbar c$ would be $197.329$~MeV-fm. \\
\textbf{\underline{Region of Interest:}} The potential has a certain range over which its influence is felt and dies down exponentially to zero. The limit value of distance $r_f$, at which $V(r_f)$ is zero is taken to be little greater than interaction radius of neutron-proton interaction. Typically, nuclear force saturates within $4$~fm and hence we have chosen $r_f = 5$~fm. The interval [0, 5] for $r$ is sampled uniformly with step-size ($h$) to obtain best accuracy.\\
\textbf{\underline{Choice of Numerical technique:}}
One must consider three key characteristics of stability, accuracy and efficiency, in that order of importance, while choosing numerical technique for solving any problem. Typically, $2^{nd}$ or $4^{th}$ order Runge-Kutta methods can be utilized for solving phase equation. But, for $np$-interaction, experimental SPS are known to three decimal places (See supplemental material, Table 1 in Appendix). The global error for RK-4 method is of the order $h^4$ and this could further add up due to propagation errors accrued with number of iterations. So, $5^{th}$ order Ruge-Kutta method (RK-5) is suggested for obtaining scattering phase shifts for $np$-interaction\cite{Zhaba}, even though RK-4 should suffice for implementation at undergraduate level. RK-5 is an interesting technique, that involves determination of 6 slopes even though only 5 are utilised for updating the solution at the next step.\\ 
The phase equation can be viewed as
\begin{equation}
\frac{d\delta(r)}{dr} = f(r,k,V,\delta)
\end{equation}
where
\begin{equation}
f(r,k,V,\delta) = -\frac{2\mu}{\hbar^2}\frac{V(r)}{k}sin^2(kr+\delta(r))
\end{equation}
The method involves calculating value of $\delta(r_{i+1})$ by utilizing previous value at $r_i$, for i = 0, 1,$\ldots$, n-1. 
\begin{equation}
\delta(r_{i+1}) = \delta(r_{i}) + \frac{h}{90}(7F_1 + 32F_2 + 12 F_4 + 32F_5 + 7F_6)
\end{equation}
where $F_i's$ are slopes of function $f$ at different points in interval [0, h]. These function evaluations become evident on looking at algorithm presented during implementation stage. 
\subsection{Implementation of the numerical method in a computer}
\noindent{This stage involves writing an algorithm or pseudo code as it's first step. Broad blocks of algorithm are identified as:\\ \textbf{1.} Initialisation \textbf{2.} Potential Definition \textbf{3.} Function Definition \\ \textbf{4.} RK-5 procedure \textbf{5.} Outputs.} \\
The Scilab code for determining scattering phase shifts using RK-5 method has been given in Github\cite{Github}, which clearly delineates all steps given above.\\
\textbf{Optimizing model parameters:}\\
Experimental data for scattering phase shifts have been modified due to newer inputs at extended lab energies and also better accuracies at existing energies. So, one is left with a challenge of obtaining new set of model parameters that match with experimental data. There are many optimization algorithms\cite{RANSAC} for obtaining best model parameters for a chosen system. We have given our implementation\cite{Anil,VMC,Swapna} based on Variation Monte Carlo technique on Github\cite{Github}.
\subsection{Experimental Observables:}
\noindent{\textbf{Scattering Properties:}}\\
For low energy scattering, scattering length `$a$' and effective range `$r_0$' can be calculated by using relation \cite{Darewych}
\begin{equation}
k \cot(\delta)=-\frac{1}{a} + 0.5 r_0 k^2
\label{reg}
\end{equation}
Using scattering phase shifts, obtained by numerically solving phase equation, $k \cot(\delta)$ was plotted as a function of $0.5k^2$. This results in a straight line and scattering parameters $r_0$ and $a$ are obtained from its slope and intercept respectively.\\
\textbf{Partial and Total Cross section:}\\
The partial cross section for 
S-wave (i.e $\ell=0$ partial wave), is given by\cite{Krane}:
  \begin{equation}
     \sigma(k)=\frac{4\pi}{k^2}\sin^2{\delta} 
     \label{cross1}
 \end{equation}
Using numerically obtained scattering phase shifts for triplet $^3S_1$ and singlet $^1S_0$ states, we compute respective scattering cross section $\sigma_t$ and $\sigma_s$.\\
The total cross section of S-wave\cite{Krane, Bertulani} $np$ scattering is calculated as
 \begin{equation}
     \sigma(k)=\frac{3}{4}\sigma_t+\frac{1}{4}\sigma_s 
     \label{cross2}
 \end{equation} 
\section{Results and Discussions}
\begin{figure}[htp]%
    \centering
    \subfloat{{\includegraphics[height = 7cm,width=7.7cm]{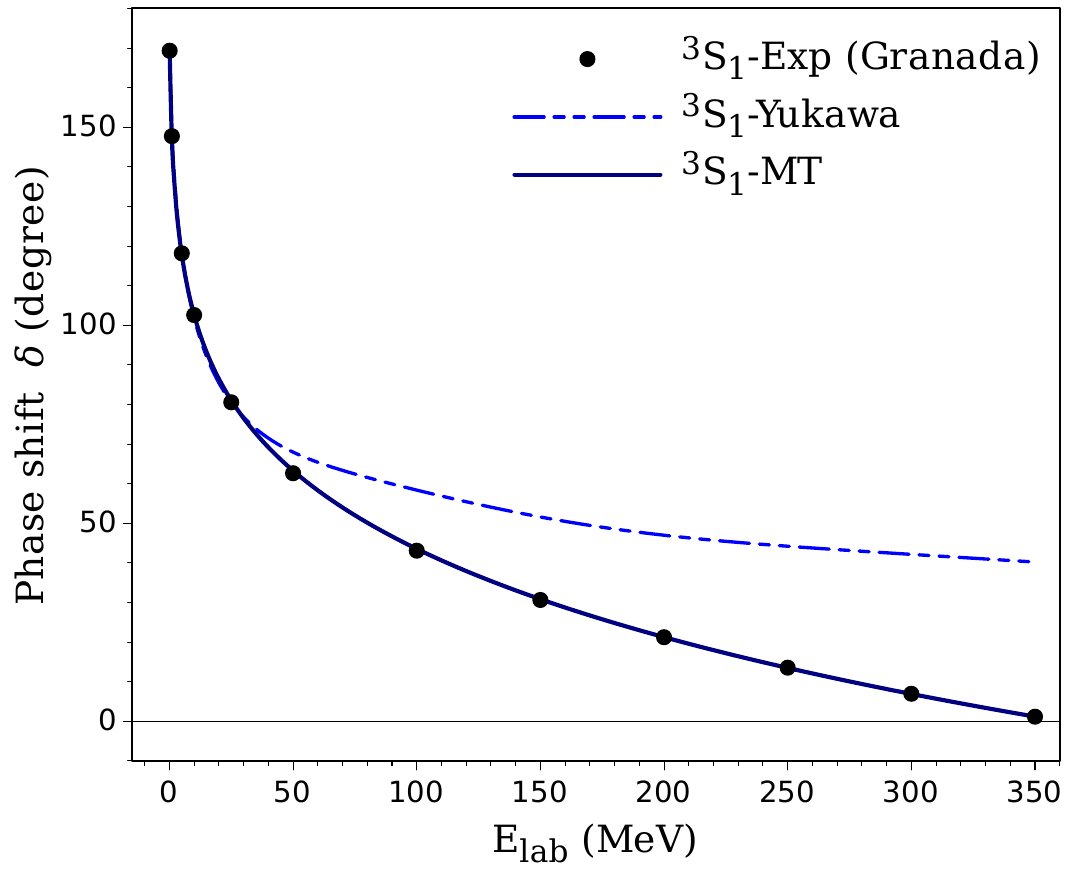} }}%
    \qquad
    \subfloat{{\includegraphics[height = 7.12cm, width=7.72cm]{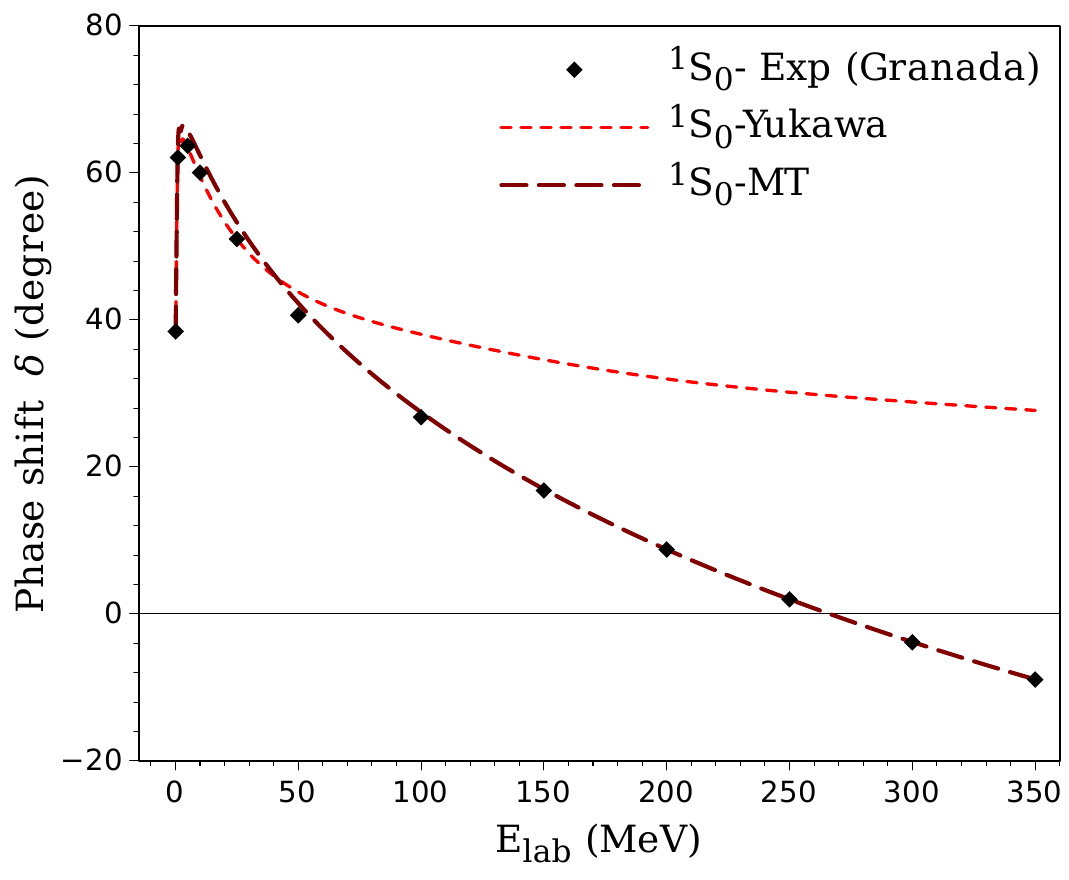} }}%
    \caption{Scattering phase shifts for (a) triplet $^3S_1$ \textit{(left)} and (b) singlet $^1S_0$ \textit{(right)} state of \textit{np} scattering obtained using Yukawa and Malfliet-Tjon potentials, along with experimental data \cite{Granada}, for lab energies upto $350$~MeV.}
    \label{Yukawa}%
\end{figure}

\begin{figure}[htp]%
    \centering
    \includegraphics[scale=0.27]{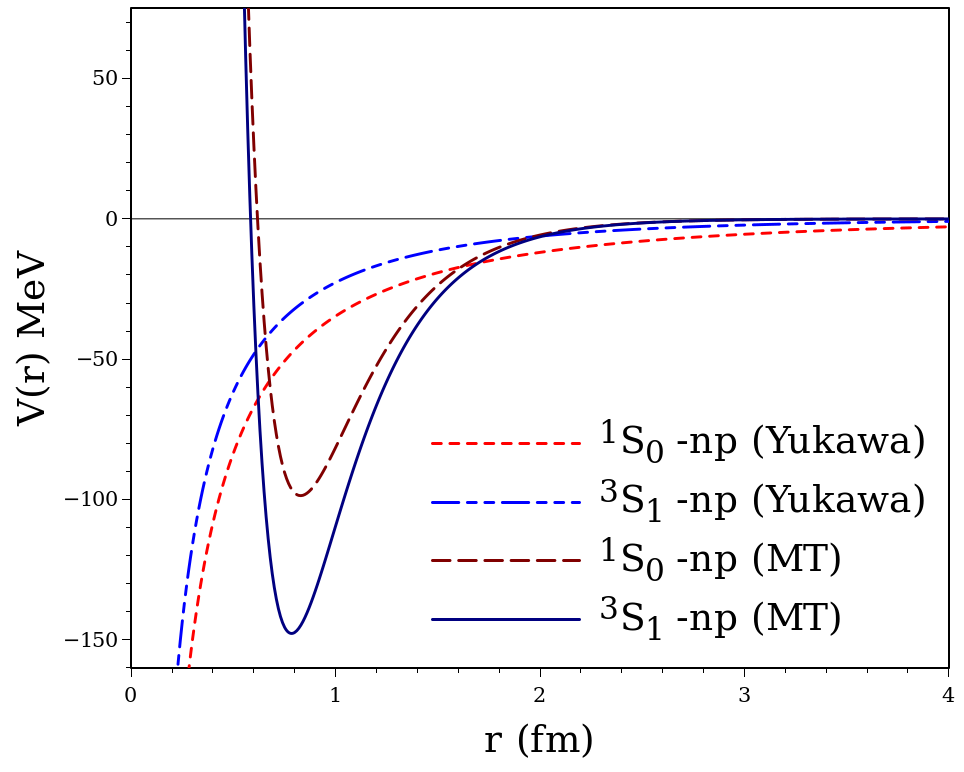}
    \caption {Plots of Yukawa and MT potentials responsible for scattering phase shifts observed due to scattering from $^3S_1$ and $^1S_0$ states in \textit{np} interaction.}
    \label{potential2}%
\end{figure}

\begin{figure}[htp]%
    \centering
    \includegraphics[scale=0.50]{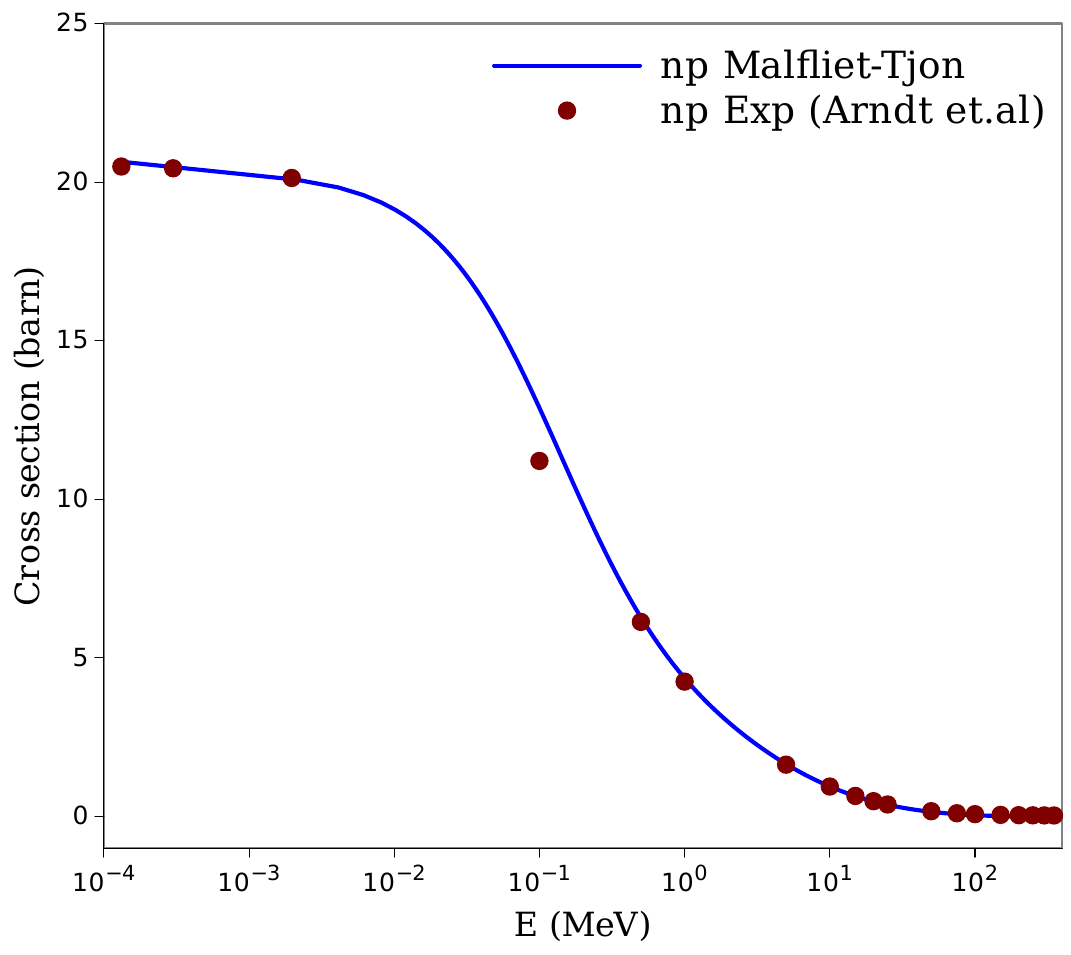}
    \caption{Neutron-proton $(np)$ experimental\cite{Arndt} and calculated cross section up to $E =350$~MeV.}
    \label{crossec}%
\end{figure}

\noindent{Experimental SPS from R. N. Pérez $et.al.$ (Granada group)\cite{Granada} along with $0.1$~MeV data point from Nijmegen database\cite{NN} have been considered for both $^3S_1$ and $^1S_0$ states.} Model parameters for both Yukawa and MT interaction potentials have been obtained based on an optimization procedure \cite{Anil} given at Github\cite{Github}, and are given in Table \ref{parameters}. The corresponding mean absolute percentage error (MAPE) are also given in Table \ref{parameters}. Utilizing these model parameters for Yukawa and MT potentials, phase equation Eq. \ref{pheqn} has been solved using RK-2, RK-4 and RK-5 methods, for both triplet $^3S_1$ and singlet $^1S_0$ states. Even though mean absolute percentage error for RK-5 and RK-4 methods are exactly similar, for two data points there has been a change in phase shift values at the $3^{rd}$ decimal place. Hence, one can implement RK-4 method in the lab, as it is well known algorithm.\\
Obtained scattering phase shifts are plotted in Fig. \ref{Yukawa}. The scattering phase shifts using RK-4 method are similar to RK-5 method. Even in RK-2 method, the change in scattering phase shifts occurs in second decimal place, so they would not be visible in plots, and hence not included.\\
\begin{table}[h!]
\centering
\caption{Model parameters and mean absolute percentage error (MAPE) for triplet ($^3S_1$) and singlet ($^1S_0$) states of \textit{np} interaction  for Yukawa and Malfliet-Tjon potentials. For Yukawa potential, the MAPE is given up-to $50$~Mev since, beyond that the MAPE value increases to a significant amount.}
\begin{ruledtabular}
\begin{tabular}{l c c c c c}
Potential & States & $V_r$~(MeV) & $V_a$~(MeV) & $\mu_A$~($fm^{-1}$) & MAPE $\%$\\
\hline
\multirow{2}{*}{Yukawa}&$^3S_1$ & --- & \ \ 50.25 & 0.37 & 1.65 (up-to $50$~MeV)\\
&$^1S_0$ & --- & \ \ 41.79 & 0.61 & 1.52 (up-to $50$~MeV)\\
\hline	
\multirow{2}{*}{Malfliet-Tjon} &$^3S_1$ & 9435.57 & \ \ 2134.88 & 2.54 & 0.61 (up-to 350~MeV)\\
&$^1S_0$ & 6806.60 & \ \ 1522.42 & 2.42 & 1.91 (up-to $350$~MeV)\\
\end{tabular}
\end{ruledtabular}
\label{parameters}
\end{table}
It should be emphasised that while CoM energies are used for computing scattering phase shifts, plots are made with laboratory energies for ease of comparison with experimental data. It is seen that scattering phase shifts for both $^3S_1$ and $^1S_0$ states, obtained using Yukawa potential match with empirical data for laboratory energies up to about $50$~MeV. Then, they start to deviate and tend to saturate to values far above expected ones.\\
On the other hand, scattering phase shifts obtained on solving phase equation using MT potential match expected data all the way upto $350$~MeV. Close match between computed and and experimental scattering phase shifts for both $^3S_1$ and $^1S_0$ states using MT potential with RK-5, RK-4 and RK-2 methods can further be observed from data compiled (See supplemental material, Table 1 in Appendix).\\
Plots of Yukawa and MT potentials using model parameters are shown in Fig. \ref{potential2}. It is interesting to observe that using MT potential, shapes of both triplet ground state and singlet scattering state are very similar except for their depths.\\ 
In order to determine low energy scattering parameters using Eq.\ref{reg}, we have considered energies from $0.1-10$~MeV which results in $k$ values from $0.035-0.347$~$fm^{-1}$. Considering scattering phase shifts obtained using MT potential, we have plotted $k cot(\delta)$ w.r.t. $0.5k^2$ for both singlet and triplet states. These are shown in Fig. \ref{reg}. Slopes of their regression lines give scattering length `$a$' and intercepts are utilized for determining effective range `$r_0$'. Obtained values are tabulated alongside experimental ones in Table \ref{scat}. While there is a good overlap between ranges of calculated experimental values\cite{Darewych} for scattering length `$a$', those for effective range `$r_0$' are reasonably close. Adding more low energy data points below $0.1$~MeV might further improve determination of these parameters.\\
\begin{figure}[htp]%
    \centering   
    \includegraphics[scale=0.8]{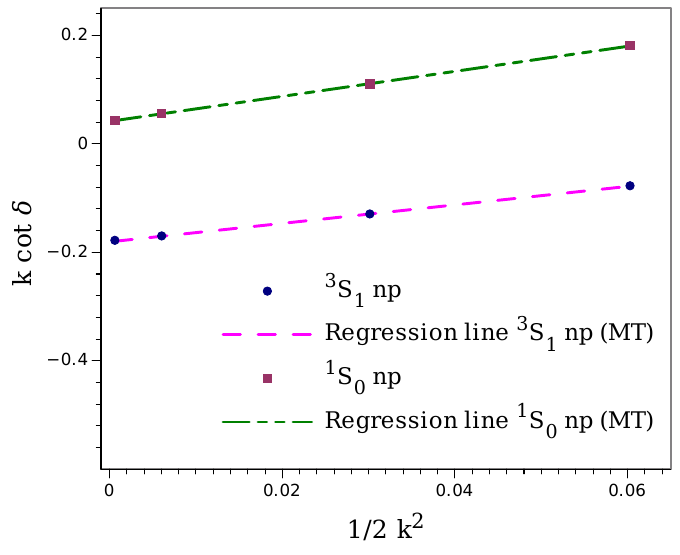}
    \caption{Plots of $k~ cot \delta$ with $0.5 k^2$ for triplet and singlet states along with regression lines}
    \label{reg}%
\end{figure}
Finally, partial and total scattering cross section were calculated from obtained scattering phase shifts using Eqns. \ref{cross1} and \ref{cross2} respectively, for experimental energies ranging from $0.1-350$~MeV. Obtained total cross section is plotted along with experimental data \cite{Arndt} in Fig. \ref{crossec}. On extrapolating to $E = 0.000132$~MeV, total cross section for S-wave is calculated to be $=20.641$~b. This is in good agreement with experimental total deuteron cross section value of $20.491$~b\cite{Krane}.\\
\begin{table}[h!]
\centering
\caption{Comparison of obtained scattering length `$a$' and effective range `$r_0$' with experimental values:}
\begin{ruledtabular}
\begin{tabular}{l c c c c c }
States & $a(fm)$ (exp.)\cite{Darewych} & $a(fm)$ (calc.) & $r_0(fm)$ (exp.)\cite{Darewych} & $r_0(fm)$ (calc.)\\
\hline	
$^3S_1$ & 5.397 $\pm$ 0.011 & 5.534 $\pm$ 0.032 & \ \ 1.727 $\pm$ 0.013 & 1.705 $\pm$ 0.012\\
$^1S_0$ & -23.678 $\pm$ 0.028 & -24.038 $\pm$ 0.045 & \ \ 2.44 $\pm$ 0.11 & 2.307 $\pm$ 0.025\\
\end{tabular}
\end{ruledtabular}
\label{scat}
\end{table}
Generally, interaction potentials that explain experimental scattering phase shifts of $np$-scattering are utilized to determine properties of deuteron which is a weakly bound nucleus consisting of one neutron and one proton. By solving the radial time-independent Schr$\ddot{\text{o}}$dinger equation using numerical technique\cite{AJPaditi}, we have determined deuteron binding energy (BE) = $-2.026$~MeV for obtained $^3S_1$ potential. The experimental binding energy of Deuteron $= -2.225$~MeV\cite{Krane}. Similarly, utilizing potential for $^1S_0$, an energy value of $76$~KeV has been obtained for unbound state which is very close to expected value of $77$~KeV. This type of determination of certain property of a system which consists of projectile and target particles is called off-shell calculation and acts as a cross-confirmation of obtained interaction potential to be consistent with other expected data. \\
The proposed method for determination of scattering phase shifts for $np$ scattering is also applicable to study $neutron-deuteron$ ($nd$)\cite{Triton} and $neutron-alpha$ ($n-\alpha$)\cite{Lalit} scattering systems. One has to redetermine model parameters for chosen interaction potential by fitting simulated scattering phase shifts to match with available experimental data for these systems. Since by changing interacting particles, few physical observables i.e, reduced mass of the system, interaction potential between interacting particles and scattering phase shifts will change. Other two body scattering systems involving charged particles, such as $proton-proton$ ($pp$)\cite{Anil} and $\alpha-\alpha$\cite{Alpha} require introducing screened Coulomb potential such as Hulthen potential, a modified form of Yukawa potential.
\section{Conclusion}
\noindent{The advantage of phase function method (PFM), which directly allows determination of scattering phase shifts without recourse to wavefunction, has been utilized to introduce phase wave analysis procedure to obtain scattering cross section.}
Phase equation for $\ell = 0 $, derived from the radial time-independent Schr$\ddot{\text{o}}$dinger equation, has been numerically solved using $5^{th}$ order Runge-Kutta (RK-5) method for determining scattering phase shifts of both triplet and singlet states of $np$ - interaction.\\ 
While attractive Yukawa potential was able to explain experimental scattering phase shifts data for lab energies upto about $50$~MeV, MT potential consisting of an extra repulsive part performed well even at higher energies upto $350$~MeV. These scattering phase shifts were utilized to obtain scattering length and effective range for both S-waves and are found to be having good match with experimental values.\\
Finally, total scattering cross section at various energies have been obtained from partial scattering cross section of both S-waves to very good accuracy. The methodology detailed in this paper could be easily extended to study other two body scattering phenomenon in atomic, nuclear and particle physics and hopefully is within reach of undergraduate physics students to undertake interesting projects. \\
 
\textbf{Acknowledgement:} We are very thankful to Prof. C. Rangacharyulu, from University of Saskatchewan, for his inputs in improving presentation of this paper.

\textbf{Conflict of Interest:} The authors have no conflicts to disclose.


\begin{thebibliography}{99}
\bibitem{Yukawa} H. Yukawa, S. Sakata, M. Kobayasi, and M. Taketani. ``On the Interaction of Elementary Particles. IV." Progress of Theoretical Physics Supplement \textbf{1}, 46-71 (1955).
\bibitem{quantYuk} M. Napsuciale, and S. Rodríguez. ``Complete analytical solution to the quantum Yukawa potential." Physics Letters B \textbf{816}, 136218 (2021).
\bibitem{Krane} K. S. Krane, \textit{Introductory nuclear physics}. (John Wiley $\&$ Sons, 1991).
\bibitem{Wong} S. S. M. Wong, \textit{Introductory nuclear physics}. (John Wiley $\&$ Sons, 2008).
\bibitem{Hans} H. S. Hans,  \textit{Nuclear Physics: experimental and theoretical}. (New Age International, 2008).
\bibitem{MT} R. A. Malfliet, and J. A. Tjon. ``Solution of the Faddeev equations for the triton problem using local two-particle interactions." Nuclear Physics A \textbf{127} (1), 161-168 (1969).
\bibitem{rmatrix} E. P. Wigner, and L. Eisenbud. ``Higher angular momenta and long range interaction in resonance reactions." Physical Review \textbf{72} (1), 29 (1947).
\bibitem{smatrix} R. S. Mackintosh,  ``Inverse scattering: applications to nuclear physics." arXiv preprint arXiv:1205.0468 (2012).
\bibitem{jost} R. Jost, and A. Pais. ``On the scattering of a particle by a static potential." Physical review \textbf{82} (6), 840 (1951).
\bibitem{Thijssen} J. Thijssen, \textit{Computational Physics}. 2nd ed. Cambridge: (Cambridge University Press, 2007). doi:10.1017/CBO9781139171397.
\bibitem{Anil}  O. S. K. S. Sastri, A. Khachi, and L. Kumar. ``An Innovative Approach to Construct Inverse Potentials Using Variational Monte-Carlo and Phase Function Method: Application to $np$ and $pp$ Scattering." Brazilian Journal of Physics \textbf{52} (2), 1-6 (2022).
\bibitem{Triton}  S. Awasthi, A. Khachi, L. Kumar and O. S. K. S. Sastri. ``Triton scattering phase-shifts for S-wave using Morse potential." Journal of Nuclear Physics, Material Sciences, Radiation and Applications \textbf{9}, no. 1, 81-85 (2021).
\bibitem{Lalit}  L. Kumar, S. Awasthi, A. Khachi and O. S. K. S. Sastri. ``Phase Shift Analysis of Light Nucleon-Nucleus Elastic Scattering using Reference Potential Approach." arXiv preprint arXiv:2209.00951 (2022).
\bibitem{Alpha} A. Khachi,  L. Kumar and O. S. K. S. Sastri. ``Alpha–Alpha Scattering Potentials for Various $\ell$-Channels Using Phase Function Method." Physics of Atomic Nuclei \textbf{85}, no. 4, 382-391 (2022).
\bibitem{Morse} P. M. Morse, ``Diatomic molecules according to the wave mechanics. II. Vibrational levels." Physical review \textbf{34} (1), 57 (1929).
\bibitem{Babikov}  V. V. Babikov, ``The phase-function method in quantum mechanics.” Soviet Physics Uspekhi \textbf{10} (3), 271 (1967).
\bibitem{Calogero} F. Calogero, ``Variable Phase Approach to Potential Scattering by F Calogero." Elsevier, \textbf{1967}.
\bibitem{Kahn} A. H. Kahn, ``Phase-Shift method for one-dimensional scattering." Am. J. Phys. \textbf{29}, 77 (1961).
\bibitem{AnilChitkara} A. Khachi, L. Kumar and O. S. K. S. Sastri.``Neutron-Proton Scattering Phase Shifts in S-Channel using Phase Function Method for Various Two Term Potentials." Journal of Nuclear Physics, Material Sciences, Radiation and Applications \textbf{9}, no. 1, 87-93 (2021).
\bibitem{Zhaba} V. Zhaba.``Calculation of phases of $np$- scattering up to Tlab=3 GeV for Reid68 and Reid93 potentials on the phase-function method."arXiv preprint arXiv:1604.06006 (2016).
\bibitem{Laha} J. Bhoi and U. Laha. ``Nucleon-Nucleon Scattering Phase Shifts via Supersymmetry
and the Phase Function Method."  Brazilian Journal of Physics \textbf{46}, 129-132 (2016).
\bibitem{rk5}  J. C. Butcher, ``On fifth order Runge-Kutta methods." BIT Numerical Mathematics \textbf{35} (2), 202-209 (1995).
\bibitem{Aditi1} O. S. K. S. Sastri, A. Sharma, S. Awasthi, A. Kachi, and L. Kumar. ``Simulation of vibrational spectrum of diatomic molecules using Morse potential by matrix methods in gnumeric worksheet." Phys. Educ. \textbf{36}, 1-14 (2019).
\bibitem{AJP} A. Sharma, S. Gora, J. Bhagavathi, and O. S. K. Sastri. ``Simulation study of nuclear shell model using sine basis." Am. J. Phys. \textbf{88}, 7, 576-585 (2020).
\bibitem{Sarsour} M. Sarsour, T. Peterson, M. Planinic, S.E. Vigdor, C. Allgower, B. Bergenwall, J. Blomgren,  T. Hossbach, W.W. Jacobs, C. Johansson and J. Klug. ``Measurement of the absolute differential cross section for $np$ elastic scattering at 194 MeV." Physical Review C, \textbf{74} (4), 044003 (2006).
\bibitem{Zettli}  N. Zettili, 2009. \textit{Quantum mechanics: concepts and applications}. (John Wiley $\&$ Sons, 2009).
\bibitem{Paneru} S.N. Paneru, ``Elastic Scattering of $^3He$ + $^4He$ with SONIK." Ohio University, (2020).
\bibitem{MIT notes} Lecture notes on:``Applied Nuclear Physics'', MIT OpenCourseWare, Massachusetts Institute of Technology. \href{https://ocw.mit.edu/courses/22-101-applied-nuclear-physics-fall-2003/pages/lecture-notes/}{(https://ocw.mit.edu/courses/22-101-applied-nuclear-physics-fall-2003/pages/lecture-notes/)}.
\bibitem{Bertulani} C. A. Bertulani and Pawel Danielewicz, \textit{Introduction to nuclear reactions}. (pp. 47-48), (CRC Press, 2019).
\bibitem{Morse2} P.M. Morse, and W.P. Allis. ``The effect of exchange on the scattering of slow electrons from atoms." Physical Review, \textbf{44} (4), 269 (1933).
\bibitem{Github} \href{https://github.com/Anynomous123/Phase-Function-Method-for-Neutron-Proton-$np$-Scattering-in-Scilab-Malfliet-Tjon-Yukawa-Potential}{https://github.com/Anynomous123/Phase-Function-Method-for-Neutron-Proton-$np$-Scattering-in-Scilab-Malfliet-Tjon-Yukawa-Potential}.
\bibitem{RANSAC}  M. A. Fischler, and R. C. Bolles. ``Random sample consensus: a paradigm for model fitting with applications to image analysis and automated cartography." Communications of the ACM \textbf{24}, no. 6, 381-395 (1981).
\bibitem{VMC}  A. Sharma and O. S. K. S. Sastri. ``Numerical solution of Schr$\ddot{o}$dinger equation for rotating Morse potential using matrix methods with Fourier sine basis and optimization using variational Monte‐Carlo approach." International Journal of Quantum Chemistry \textbf{121} (16), e26682 (2021).
\bibitem{Swapna} S. Gora, O. S. K. S. Sastri, and S. K. Soni. ``Optimization of semi-empirical mass formula co-efficients using least square minimization and variational Monte-Carlo approaches." European Journal of Physics \textbf{43} (3), 035802 (2022).
\bibitem{Darewych} G. Darewych, and A. E. S. Green. ``Morse Function and Velocity-Dependent Nucleon-Nucleon Potentials." Physical Review \textbf{164}, 1324 (1967).
\bibitem{Granada} R. N. Pérez, J. E. Amaro, and E. Ruiz Arriola. ``The low-energy structure of the nucleon–nucleon interaction: statistical versus systematic uncertainties." Journal of Physics G: Nuclear and Particle Physics \textbf{43} (11), 114001 (2016).
\bibitem{NN} NN-Online (\href{https://nn-online.org/NN/}{https://nn-online.org/NN/}).
\bibitem{Arndt} R. A. Arndt, W. J. Briscoe, A. B. Laptev, I. I. Strakovsky, and R. L. Workman. ``Absolute Total $np$ and $pp$ Cross Section Determinations." Nuclear science and engineering \textbf{162} (3), 312-318 (2009).
\bibitem{AJPaditi} A. Sharma, S. Gora, J Bhagavathi, and O. S. K. S. Sastri. ``Simulation study of nuclear shell model using sine basis." Am. J. Phys. 88, \textbf{7}, 576-585 (2020).
\end{thebibliography}
\end{document}



\title{Numerical Simulation Study of Neutron-Proton Scattering using Phase Function Method}

\author{Shikha Awasthi}
\author{Anil Khachi}
\author{Lalit Kumar}
\author{O.S.K.S. Sastri}
\email{sastri.osks@hpcu.ac.in} 
\affiliation{Department of Physics and Astronomical Science, Central University of Himachal Pradesh, Dharamshala - 176215, Bharat (India)}

\date{\today}

\maketitle 

\section{Appendix 1}
Experimental and obtained SPS obtained from RK-5, RK-4 and RK-2 methods for $^3S_1$ and $^1S_0$ states, using MT potential, are given below in Table \ref{nppp1} and \ref{nppp2}. Mean absolute percentage error (MAPE) for triplet state are determined to be 0.61$\%$, 0.61$\%$ and 0.63$\%$ respectively.\\
\begin{table}[h!]
\centering
\caption{Experimental\cite{Granada} and obtained SPS, for triplet $^3S_1$ state of \textit{np} interaction for 0.1-350 MeV lab energies}
\begin{ruledtabular}
\begin{tabular}{l c c c c}
$E_{lab}$ &  $^3S_1$(Exp)\cite{Granada} &  $^3S_1$(MT) RK-5 & $^3S_1$(MT) RK-4 &  $^3S_1$(MT) RK-2\\
(MeV) &  $\delta_{np}$ &  $\delta_{np}$ &  $\delta_{np}$ &  $\delta_{np}$\\
\hline	
0.1 & 169.320 &         \ \    169.042  &  169.041   &  168.977 \\
1 &  147.748 & \ \  147.185   &  147.185     &  147.116  \\
5 &  118.169 & \ \ 117.827    &  117.827     &  1102.568  \\
25 &  80.559 & \ \  81.058    &  81.058     & 81.042  \\
50 & 62.645  & \ \  63.296    &  63.296  & 63.287 \\
100 &  43.088 & \ \  43.549   &  43.549    & 43.545  \\
150 &  30.644 & \ \  30.846   &  30.846   	& 30.844  \\
200 &  21.244 & \ \  21.264    &  21.265    &  21.264   \\
250 &  13.551 & \ \  13.494   &  13.494     &  13.494  \\
300 &  6.966  & \ \  6.919     &  6.919    & 6.919 \\
350 & 1.176  & \ \  1.199    &  1.199    &	 1.199 \\
\hline
MAPE ($\%$)&& 0.61 & 0.61 & 0.63\\
\end{tabular}
\end{ruledtabular}
\label{nppp1}
\end{table}
Mean absolute percentage error (MAPE) for singlet state using RK-5, RK-4 and RK-2 methods are determined to be 0.91$\%$, 1.91$\%$ and 1.93$\%$ respectively.\\
\begin{table}[h!]
\centering
\caption{Experimental\cite{Granada} and obtained SPS, for triplet $^3S_1$ and singlet $^1S_0$ states of \textit{np} interaction for 0.1-350 MeV lab energies}
\begin{ruledtabular}
\begin{tabular}{l c c c c}
$E_{lab}$ &  $^1S_0$(Exp)\cite{Granada} &  $^1S_0$(MT) RK-5 & $^1S_0$(MT) RK-4 &  $^1S_0$(MT) RK-2\\
(MeV) &  $\delta_{np}$ &  $\delta_{np}$ &  $\delta_{np}$ &  $\delta_{np}$\\
\hline	
0.1 &  38.43   &  38.42 &  38.42   &  38.33 \\
1 &    62.105     &  63.018  &  63.018     &  62.962 \\
5 &   63.689     &  65.725  &  65.725     &  65.705 \\
10 &  60.038    &  62.412  & 62.412    &  62.400 \\
25 &   51.011     & 53.283 & 53.283     & 53.278 \\
50 &   40.644   & 42.284 &  42.284   & 42.282 \\
100 &   26.772    &  27.444  &  27.444    &  27.444 \\
150 &  16.791   	&  16.975  &  16.975   	&  16.976 \\
200 &   8.759    &  8.777   &  8.777    &  8.778 \\
250 &  1.982     &  1.995  &  1.995     &  1.996  \\
300 &   -3.855    & -3.812 &  -3.812    & -3.811 \\
350 & -8.923    &	 -8.903 &  -8.903    &	-8.901 \\
\hline
MAPE ($\%$)&& 1.91 & 1.91 & 1.93\\
\end{tabular}
\end{ruledtabular}
\label{nppp2}
\end{table}
\newpage
\section{Appendix 2}
\textbf{Scilab code for determining scattering phase shifts of $^3S_1$ state of $np$ interaction using $5^{th}$ order Runge-Kutta method:}\\
\textbf{Step 1: Initialisation}
\begin{verbatim}
//Set values for object variables:
mp = 938.272046;                // mass of proton in MeV/c^2
mn = 939.565379;               //mass of neutron in MeV/c^2
Elab = [0.1, 1, 5, 10, 25, 50, 100, 150, 200, 250, 300, 350]
                               // Experimental lab energy
ObjVar = [mn,mp,Elab];
//Set values for interaction variables:
Vr = 0; Va = 50.25;  muA = 0.37;//Yukawa model parameters for 3S1
// Vr = 0; Va = 41.79;  muA = 0.61;Yukawa model parameters for 1S0
// Vr = 9435.72; Va = 2134.90;  muA = 2.54; MT model parameters for 3S1
// Vr = 6806.60; Va = 1522.42;  muA = 2.42; MT model parameters for 1S0
 
Vparam = [Vr, Va, muA];
//Set values for State variables:
deltaExpt = [169.32, 147.748, 118.169, 102.587, 80.559, 62.645, 43.088, 30.644, 21.244, 13.551, 6.966, 1.176] // Exp. SPS for 3S1
//deltaExpt = [38.43, 62.105, 63.689, 60.038, 51.011, 40.644, 26.772, 16.791, 8.759, 1.982, -3.855, -8.923] // Exp. SPS for 1S0              
function [k, fac] = input(ObjVar)
    muD = (mn*mp)/(mn + mp);     // reduced mass of np system
    hbarc = 197.329              //hcut*c in units MeV fm
    fac = hbarc^2/(2*muD)  //(= 41.47) denomenator term used in function 'F'
    ECoM = Elab*mp/(mn+mp)       // converting Elab to ECoM
    k = sqrt(ECoM/fac)
endfunction
\end{verbatim}
\textbf{Step 2: Potential definition}
\begin{verbatim}
function V = Potential(r, Vparam)
    Vr = Vparam(1); Va = Vparam(2); muA = Vparam(3); 
    V = -Va.*exp(-muA.*r)./r + Vr.*exp(-2.*muA.*r)./r
endfunction
\end{verbatim}
\textbf{Step 3: Function definition}
\begin{verbatim}
function F = f(r,delta0,Vparam)
    [k,fac] = input(ObjVar)
    V = Potential(r, Vparam)
    F = (-V.*sin(k.*r + delta0).^2)./(k.*fac)
endfunction
\end{verbatim}
\textbf{Step 4: PFM using RK-5 procedure}
\begin{verbatim}
function deltaObt = PFMRK5(Vparam, deltaExpt)
    h=0.01; r0=0.01; rf=5; delta0 = 0
    for r = r0:h:rf                           // defining range of potential
        F1 = f(r,delta0,Vparam)              // RK-steps
        F2 = f(r+h/4,delta0 + h*F1/4,Vparam)
        F3 = f(r+h/4,delta0 + h*F1/8 + h*F2/8,Vparam)
        F4 = f(r+h/2,delta0 -h*F2/2 + h*F3,Vparam)
        F5 = f(r+3*h/4,delta0+3*h*F1/16+9*h*F4/16,Vparam)
        F6 = f(r+h,delta0-3*h*F1/7+2*h*F2/7+12*h*F3/7-12*h*F4/7+8*h*F5/7,Vparam)
        delta1 = delta0 + h*(7*F1+32*F3+12*F4+32*F5+7*F6)/90
        r=r+h  //incrementing distance
        delta0 = delta1                      // SPS in radian
    end
    deltaObt = delta0.*180./%pi          //converting SPS from radian to degree
\end{verbatim}
\textbf{Step 5: Outputs}
\begin{verbatim}
plot(ECoM,deltaExpt,`*',ECoM,deltaObt)   //Plotting lab energy w.r.t SPS
endfunction
\end{verbatim}
Save file as `PFMRK5.sci' and execute in Scilab. This code is also available on GitHub\cite{Github}.